\pdfoutput=1

\documentclass[conference]{IEEEtran}
\usepackage[numbers]{natbib}

\usepackage{graphicx}
\graphicspath{{images/}}

\ifCLASSINFOpdf
\else
\fi
\begin{document}
%
\title{A Debt-Aware Learning Approach for Resource Adaptations in Cloud Elasticity Management}

\author{\IEEEauthorblockN{Carlos Mera-G\'omez\IEEEauthorrefmark{1}\IEEEauthorrefmark{2}, Francisco Ram\'irez\IEEEauthorrefmark{1},  
Rami Bahsoon\IEEEauthorrefmark{1} and
Rajkumar Buyya\IEEEauthorrefmark{3}, 
}
\IEEEauthorblockA{\IEEEauthorrefmark{1}School of Computer Science\\
University of Birmingham,
Edgbaston, UK, B15 2TT\\ Email: \{cxm523, fmr067, r.bahsoon\} @cs.bham.ac.uk}
\IEEEauthorblockA{\IEEEauthorrefmark{2} ​Escuela Superior Polit\'ecnica del Litoral, ESPOL, \\ Facultad de Ingenier\'ia en Electricidad y Computaci\'on, \\Campus Gustavo Galindo Km 30.5 V\'ia Perimetral, P.O. Box 09-01-5863, Guayaquil, Ecuador\\
Email: cjmera@espol.edu.ec}
\IEEEauthorblockA{\IEEEauthorrefmark{3}
Cloud Computing and Distributed Systems (CLOUDS) Lab\\
School of Computing and Information Systems\\
The University of Melbourne, Australia\\
Email: rbuyya@unimelb.edu.au}
}


%


\maketitle

\begin{abstract}
Elasticity is a cloud property that enables applications and its execution systems to dynamically acquire and release shared computational resources on demand. Moreover, it unfolds the advantage of economies of scale in the cloud through a drop in the average costs of these shared resources. 
However, it is still an open challenge to achieve a perfect match between resource demand and provision in autonomous elasticity management. Resource adaptation decisions essentially involve a trade-off between economics and performance, which produces a gap between the ideal and actual resource provisioning. This gap, if not properly managed, can negatively impact the aggregate utility of a cloud customer in the long run. 
To address this limitation, we propose a technical debt-aware learning approach for autonomous elasticity management based on a reinforcement learning of elasticity debts in resource provisioning; the adaptation pursues strategic decisions that trades off economics against performance. 
We extend CloudSim and Burlap to evaluate our approach. The evaluation shows that a reinforcement learning of technical debts in elasticity obtains a higher utility for a cloud customer, while conforming expected levels of performance.

\end{abstract}


%
\IEEEpeerreviewmaketitle

\section{Introduction}
Elasticity is the essential characteristic of cloud computing that supports an on-demand provision and release of shared resources to meet an expected quality of service \cite{buyya2009cloud, mell2011nist}. This characteristic is one of the enablers for the cloud \textit{economies of scale} \cite{sullivan2003economics}, dropping the average  cost of computing resources \cite{armbrust2010view}. Therefore, elasticity decisions on resource adaptation should be driven not only by performance considerations but also by an economics perspective to pursue a long-term utility under uncertainty. The uncertainty  comes from diverse sources due to likely changes in quality of service attributes (e.g. throughput, reliability, availability), dynamic budget constraints \cite{han2015investigations}, workload deviations or a \textit{resource contention} \cite{li2010cloudcmp, gambi2013testing} among others.

Although elasticity continuously performs dynamic resource adaptations; in practical terms, it is impossible to achieve a perfect match between resource provisioning and demand between consecutive adaptations  \cite{schulz2013elasticity, herbst2015bungee, islam2012consumer}. Therefore, this gap between the ideal and actual resource provisioning calls for a dynamic valuation that incorporates a strategic trade-off between performance and economics. On one hand, this valuation should consider that effects of elasticity adaptations on performance, for example, are not instantaneous due to the \textit{spin-up time} \cite{brebner2012your, li2010cloudcmp}. On the other hand, the same valuation should consider that the economics of these adaptations depends on  billing cycles, pricing schemes and resource bundles granularity \cite{suleiman2012understanding}; as in the case of a \textit{partial usage waste} \cite{jin2015towards}, which results from the additional time charged for a resource between its release and the end of the billing cycle.



In our previous work \cite{meraelasticity}, we proposed an elasticity conceptual model that identifies \textit{technical debts} that are linked to cloud elasticity adaptations taken under uncertainty, and we defined the term \textit{elasticity debt} as the valuation gap between the ideal and actual resource provisioning in elasticity adaptations. 

The novel contribution of this paper is an elasticity management approach that autonomously learns the value of elasticity debts and dynamically trades off performance against economics in adaptation decisions. The adaptation pursues to take decisions that maximise the long-term utility of the elastic system by incurring strategic debts. The approach contributes to the fundamentals of technical debt management, where our work is the first to transit the debt analysis from a static to a dynamic perspective through a \textit{reinforcement learning} approach to make strategic adaptation decisions. Technical debt \cite{cunningham1993wycash} is a metaphor that supports a trade-off analysis between a quick engineering decision that yields immediate benefits at the expense of compromising long-run objectives \cite{kruchten2012technical}. Elasticity adaptation can incur an elasticity debt that renders short-term benefits but compromises performance, economics or both. The debt can accumulate  if not properly valued. These debts can be retrospectively analysed in a threshold-based rules management for elasticity or dynamically learnt with a proactive perspective in a reinforcement learning based elasticity management. Reinforcement learning \cite{sutton1998reinforcement} is an approach that seeks optimality in decision-making through a continuous learning that forgoes short-term rewards to achieve higher long-term gains.


Technical debt metaphor has been applied in software architecture, software maintenance and evolution, cloud service selection among others \cite{li2015systematic}. Additionally, elasticity management based on reinforcement learning with a pure performance perspective has been already applied \cite{barrett2013applying, lorido2014review}. However, to our knowledge, our work is the first to treat the debt as a moving target that dynamically changes over time. We shared this self-adaptive perspective for technical debt in the recent Dagstuhl Seminar 16162 \cite{avgeriou2016managing}; the suggestion was well received by the technical debt community. Moreover, the contribution is the first to introduce an online learning approach for technical debt; the approach  identifies, tracks, and monitors the debt and payback strategies of adaptation decisions in the context of cloud elasticity. We evaluate the approach through a simulation tool that extends CloudSim \cite{calheiros2011cloudsim} and Burlap \cite{burlap}. The results show that a reinforcement learning of technical debts can achieve a higher aggregate utility for a service provider.

The rest of the paper is organized as follows. Section II presents the problem statement and motivates the need for an online learning of elasticity debts, while Section III describes the motivating example and assumptions. Section IV provides a detailed overview of our debt-aware learning approach and explains its components. We report the evaluation of our approach in Section V, followed by a discussion of related works in Section VI. Finally, section VII summarizes our conclusions and directions for future research.

\section{Problem Statement}
A cloud elasticity configuration adopted by a Software as a Service (SaaS) provider (cloud customer) should aim to conform an expected performance, in terms of quality for service delivery to end users (cloud consumers), and to keep their economics, in terms of operating costs, at a minimum when acquiring or releasing resources from an Infrastructure as a Service (IaaS) provider (cloud provider). In practice, it is impossible to achieve a perfect elasticity i.e. exactly match resource supply with demand \cite{schulz2013elasticity, herbst2015bungee, islam2012consumer} due to several reasons such as the difficulty to predict resource demand, coarse computing resource granularity, elapsed time between computing resources are acquired and when they are effectively ready to be used, restrictions on the number of computing resource to be acquired at once, pricing schemes granularity and billing cycles among others \cite{jamshidi2016managing, suleiman2012understanding}. Hence, elasticity decisions should optimize for a dynamic resource provision not only in terms of performance metrics but also from an economics perspective that can maximise the utility of the SaaS provider in the long run.

Currently, elasticity is analysed from a performance \cite{herbst2015bungee, vaquero2011dynamically}, cost-aware \cite{sharma2011cost, han2014enabling} or economics-driven perspective \cite{fokaefs2016economics, pandey2016hybrid, islam2012consumer}. However, none of these approaches incorporate a strategic valuation of elastic adaptations to make explicit trade-offs in the decision-making when adjusting a resource provisioning. Consequently, these myopic adaptations lead to a provision of resources that obtains short-term gains when matching the resource demand but can be suboptimal in the long-term with hidden consequences that waste resources or degrade quality of service attributes (e.g. performance, security, reliability), which results in an affectation to the aggregate utility of the cloud customer over time. 

Technical debt metaphor supports a reasoned decision-making about quick engineering decisions taken to obtain short-term benefits at the cost of introducing liabilities that compromise long-term system objectives. In dynamic environments, the utility of these decisions can be systematically learnt through a reinforcement learning approach. Reinforcement learning is a technique where a farsighted agent learns from continuous interactions with an environment how to maximize a long-term reward without any a priori knowledge. We combine this online learning with the technical debt metaphor in the context of cloud elasticity to evaluate dynamic trade-offs carried out by elastic adaptation decisions. The consideration of debt motivates a value-oriented perspective to adaptation that systematically links the consequences of these decisions with environmental uncertainty, such as unexpected workload variations, dynamic changes in quality of service or resource failures. 

We advocate that elasticity can benefit from a debt-aware learning perspective by making the elasticity debts visible, revealing the performance and economics consequences of adaptation decisions (e.g. over- or under-provisioning states) that are prone to uncertainty and therefore improving the utility achieved by a cloud stakeholder (e.g. SaaS provider) in terms of reducing penalties that relate to Service Level Agreement (SLA) violations and operating costs minimization. 

Optimization of elastic resource provisioning in the cloud calls for the development of new approaches that consider the value and utility of adaptation decisions. In this paper, we propose a new approach to reason about the dynamic adaptations in cloud elasticity and make them more profitable by learning the trade-off decisions that potentially introduce technical debts. 
 
\section{System Model}

\subsection{Motivating Example}
A multi-tenant application is a highly configurable software that allows each tenant (client), usually an organization that serves a number of users, to customize its appearance and application workflows according to their needs, which makes it appear different for each tenant but indeed all of them are sharing a single application \citep{bezemer2010multi}. Furthermore, the application owner (provider) can negotiate individual SLAs with each tenant. An example of these type of applications is an Enterprise Resource Planning (ERP).

Let us consider a globally accessed multi-tenant SaaS survey application, where tenants after subscribing to the service can design a survey, publish it and collect its results for analysis. Simultaneously, multiple surveys from different tenants run and depending on the number of participants attracted, the service workload can experience a sudden sharp of resource demand that should be handled by the service infrastructure accordingly. The service owner is a SaaS provider who processes incoming HTTP requests, from tenants and participants, on the IaaS provider infrastructure where the service is deployed. 

The SaaS service provider needs to comply with SLAs signed with their clients and subsequently avoid penalties and loss of reputation. Simultaneously, the provider desires to incur in minimum costs related to virtual resources dynamically acquired from the IaaS provider. Therefore, elasticity adaptation decisions should be systematically evaluated; the evaluation decisions shall be  strategically geared in terms of the utility they can achieve. From a business perspective, the utility is determined by the profit achieved after processing incoming application requests over time.  


\section{Proposed Approach}

\subsection{Technical Debt on Elasticity}
Technical debt is a metaphor that makes visible the valuation of alternatives in a trade-off between an ideal and an actual decision making \cite{brown2010managing, guo2011portfolio}; where the debt is determined by the valuation of the gap between these two alternatives \cite{li2014architectural}. The metaphor has shown to be effective to identify, measure and monitors technical debts over time \cite{seaman2012using}. In our previous work \cite{meraelasticity}, we developed the foundations for introducing the built-in decision support of technical debt analysis into the large scale dynamic and adaptive context of cloud elasticity management. We defined \textit{elasticity technical debt} as the valuation of the gap between an optimal and an actual  adaptation decision. This debt trades off the performance to obtain with the provisioning of an elasticity adaptation against the economics of that adaptation. 

Likewise a debt in finance, an elasticity debt can be either \textit{strategic} or \textit{unintentional}. The former refers to adaptations that intend to anticipate changing conditions (e.g. workload variations) or mitigate undesired effects (e.g. spin-up time, partial usage waste); whereas the latter refers to delayed or wrong choice of adaptations (e.g. resource thrashing) as a consequence of poor considerations for uncertainty or elasticity determinants. The value of elasticity debts can be observed \textit{retrospectively} in traditional debt unaware elasticity management approaches, or \textit{proactively} in debt-aware approaches that utilise this valuation to analyse and decide elasticity adaptations. 

Different from traditional approaches, that mostly consider avoiding over- and under-provisioning states, we \textit{argue} that an elasticity debt-aware approach recognizes the fact that it is practically impossible to achieve a perfect elasticity; and makes use of this fact to explicitly reveal the potential of using this imperfection in the trade-off between economics and  performance to adjust strategically the resource provisioning and preserve the utility of the stakeholder. For example, we may intentionally delay an over-provisioning state if the next billing cycle of the resources to be released is not immediate; or if we consider that the spin-up time of launching new resources may affect the SLA performance compliance during a imminent growth in the load.

\begin{figure}[ht]
\centering
\includegraphics[width=\columnwidth]{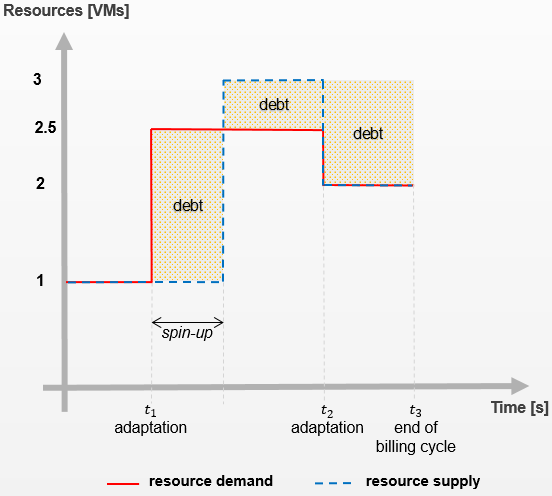}
\caption{Examples of elasticity debts}
\label{fig:debtexamples}
\end{figure}

Figure \ref{fig:debtexamples} illustrates three cases of debts using a graph that represents  a resource demand and supply over time. The first gap is caused by the spin-up time when new virtual machines are launched; the second gap is a consequence of the available resource granularity that makes impossible to launch one and a half machines; and the third less evident gap is the result of a partial usage waste after one machine is released but still charged until the end of the billing cycle. In any case, the debt is not the gap itself. We highlight that a debt corresponds to the valuation in terms of the potential utility produced by the gap, where the debt originates. 

\subsection{Reinforcement Learning}
Reinforcement learning \cite{sutton1998reinforcement} is a framework that pursues an optimal decision-making based on the maximization of a cumulative reward in the long-term. The decision-maker or \textit{agent} learns through consecutive interactions with an \textit{environment}, where each \textit{action} modifies the environmental \textit{state} and produces a \textit{reward}, which is the utility that the agent receives from the action. Both, the set of variables that characterizes the new state and the reward are perceived by the agent. This learning technique has already been applied to cloud elasticity management \cite{lorido2014review, barrett2013applying}, where an agent takes resource adaptation decisions based on the current state, which is usually identified by performance thresholds, and achieves a reward, which is given by the new performance monitored after the adaptation takes place.

We follow a model-free reinforcement learning strategy rather a model-based because our learning environment lacks of a predefined transition model that describes the effect of each action \( a \) in a given state \( s \) by determining the probability of reaching a specific subsequent state \( s_{t+1} \) \citep{russell2016artificial}. A model-free strategy uses an \textit{action-utility function}, known as \( Q(s,a) \), to estimate the value of performing an action \( a \) over a state \( s \). From the available algorithms in this kind of learning strategy, we have adopted \textit{Q-learning} \cite{sutton1998reinforcement} because it is more flexible to explore changes in the environment, making it more convenient for highly dynamic contexts \citep{dewolf2013}. Furthermore, it is the most common extended algorithm with respect to elasticity management \citep{lorido2014review}. 

The Q-learning algorithm learns an optimal decision-making by repeatedly updating the utility of an action \( a \) given a state \( s \) according to the following update rule: 

\begin{equation}
	\label{Q_eq}
 	Q(s,a) \leftarrow (1-\alpha)*Q(s,a) + \alpha*[r+\gamma*max_{a_{t+1}}Q(s_{t+1}, a_{t+1})],
\end{equation}  

where \(\alpha\) is the learning rate (a value that usually starts at 1 and decreases over time), \( r \) is the reward of the action, \(\gamma\) is the discount factor (a value between 0 and 1 that adjusts a learner from myopic to far-sighted respectively), and \( s_{t+1} \) is the resulting state, and  \( a_{t+1} \) is the best possible action to take thereafter. 

Interactions with the environment are classified as \textit{exploration} or \textit{exploitation}. The former aims to perform random actions to experience environmental changes to preclude from focus on immediate gains; whereas the latter aims to only make use of what the agent already knows. This trade-off between exploration and exploitation depends on an \(\epsilon\)-greedy policy \citep{wiering2012reinforcement}, which means that a learner exploits the best action with probability (1-\(\epsilon\)) and explores a random action with probability \(\epsilon\). 

\subsection{Approach Description}
We propose an elasticity management based on a reinforcement learning of elasticity technical debts incurred by elasticity adaptations in order to make debt informed decisions. Our debt-aware learning approach explores and learns elasticity debts over time and then uses this knowledge from previous experiences to incur in strategic debts that minimizes negative effects on aggregate utility. Making use of the  function defined in \cite{pandey2016hybrid}, the utility achieved by an IaaS cloud customer, in this case a SaaS provider, when processes a workload \textit{w}, composed of jobs or incoming requests, is calculated in terms of revenue, penalty and operating costs incurred during the monitored period (i.e. between consecutive elasticity adaptations) by means of equation \ref{utility}:%

\begin{equation}
	\label{utility}
 	U(w) = R(x) * x_s - P(x) * x_f - \sum_{i=1}^{N} C (vm_i) \int_{0}^{L} m_i(t)dt,
\end{equation}  

where \textit{R(x)} and \textit{P(x)} functions return the revenues and penalties per request, respectively; $x_s$ and $x_f$ represent the number of successful and failed requests, respectively, from workload \textit{w} with respect to defined in the SLA; and $C(vm_i)$ function returns the cost of each of the \textit{N} virtual machine (VM) types corresponding to their $m_i$ launched instances over the execution time.

We calculate the debt of each adaptation as the utility difference between the actual and the ideal resource provisioning, as shown in equation \ref{Debt_eq}:
\begin{equation}
	\label{Debt_eq}
 	Elasticity Debt \leftarrow U_{actual} - U_{ideal},
\end{equation} 

where \textit{U} represents the utility obtained by a SaaS provider as cloud customer during a monitoring period. In the best scenario, the elasticity debt would be zero when the actual resource provisioning matched the ideal one required in the period. Otherwise, it will be a negative number.

\begin{figure*}[t]
\centering
\includegraphics[width=0.65\textwidth]{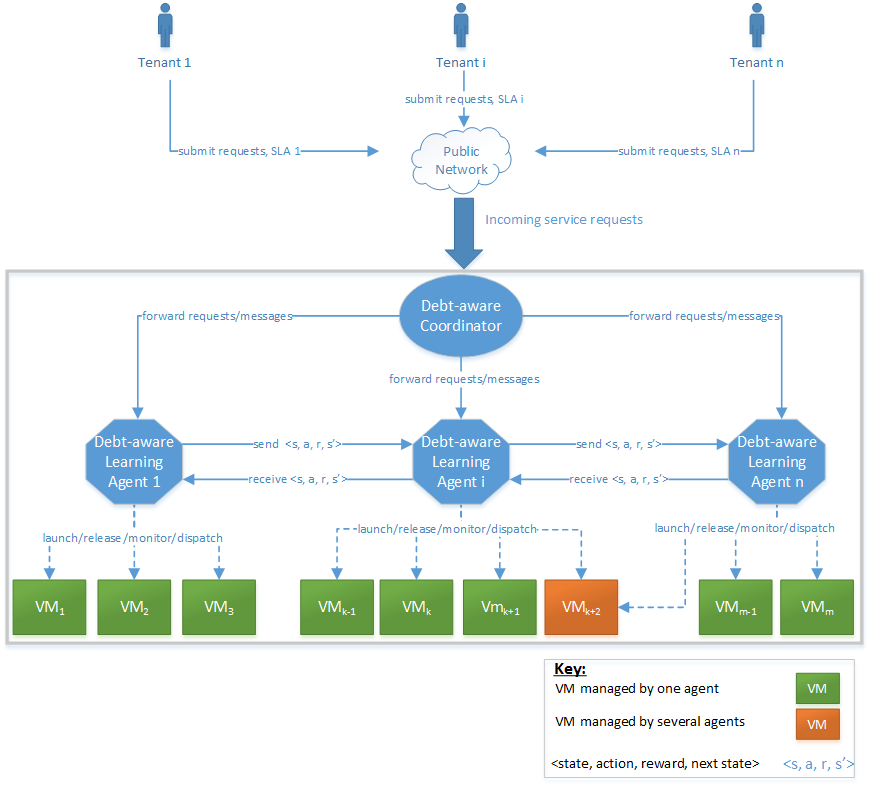}
\caption{Reference system model of our debt-aware approach}
\label{fig:model}
\end{figure*} 

A reference system model of our approach is shown in Figure \ref{fig:model}. Several tenants subscribe to a multi-tenant SaaS service with a SLA tailored to each individual need. We are assuming that the service is hosted in the infrastructure of an IaaS provider such as \textit{CloudSigma}  \cite{cloudsigma} with its pay-as-you-go pricing scheme and five minute-based billing cycle, a resource granularity in terms of VMs, and a horizontal elasticity method. These virtual resources are managed by \textit{debt-aware learning agents} that are responsible for launching, releasing, and monitoring VMs. We envision an agent-oriented architecture with hierarchy where agents tend to realise the requirements of multi-tenant users in a decentralised fashion, which promotes a scalable solution and facilitates the collaboration between different agents promising optimization for inter-agents knowledge exchange.

One of the main concerns in reinforcement learning solutions relates to the considerable training time that the algorithm takes to converge to an optimal solution \citep{lorido2014review}. To overcome this problem, we advocate a \textit{parallel reinforcement learning} mechanism \citep{mannion2015parallel}. In this mechanism, multiple agents can learn simultaneously and share their learning to speed-up the convergence time. In our model, we have grouped the running VMs in clusters and each of them is managed by a debt-aware learning agent. Furthermore, a learning agent performs a load balancing and dispatches the incoming requests, which correspond to a single tenant, to be executed in one of the VM in the cluster. A VM can be managed simultaneously by more than one learning agent to optimise  resource utilization during under-provisioned states. 

We define the elements of our reinforcement learning approach in Table \ref{table:rlelements}. A debt-aware learning agent considers the following variables to define an state: (i) a proportion of running VMs with queued request; where the proportion is categorized into high, medium or low; and (ii) a proportion of running VMs close to a next billing cycle and without queued request; where the proportion is equally categorized into high, medium or low. We avoid unnecessary exploration by including preconditions for two actions: launch and release. For instance, only launch action is available if there is a high number of VMs with queued jobs; or only  release action is permitted when a high proportion of VMs are close to a next billing cycle and without queued request. 

\begin{table}[h]
\begin{center}
\caption {Reinforcement learning elements}
  \begin{tabular}{ | l || p{0.65\columnwidth} | }
    \hline
    \textbf{Element} & \textbf{Definition} \\
    \hline
    \hline
	Environment & Cloud elasticity \\ 
	\hline
	Agent & Debt-aware learning agent, debt-aware coordinator \\ 
	\hline
	Actions & Launch, release or maintain VMs\\ 
	\hline
	State Variables & 
				\begin{enumerate}
					\item Proportion of VMs with queued requests (i.e. High, Medium and Low)
					\item Proportion of VMs close to a next billing cycle and without queued requests (i.e. High, Medium and Low)				
				\end{enumerate}					
	  \\ 
	\hline
	Reward & Elasticity Debt
	 \\ 
    \hline
  \end{tabular}  
  \label{table:rlelements}
\end{center}
\end{table}

Finally, the \textit{debt-aware coordinator} is responsible for creating and destroying learning agents, forwarding incoming service requests to the corresponding learning agent, and sending coordination messages such as changes in expected SLAs or refinements in the learning process. 

\section{Evaluation}
We compare our debt-aware reinforcement learning elasticity management against a traditional threshold-based rule one that implements the \textit{voting process} offered by \textit{Right Scale} \cite{rightscale2016}. In this voting mechanism, resource adaptations are taken based on the outcome of a voting process, where each virtual machine votes according to a performance metric (e.g. CPU utilization) decision threshold, and the agreement is reached when a specified percentage of voters supported a given decision (i.e. launch, release or maintain VMs). 

We implemented a traditional elasticity management that observes the debt in retrospective without making it to take part in the decision process. It calculates the debt of an adaptation when the next one takes place by recreating the circumstances under which the adaptation  was serving and simulating the other discarded elasticity actions to estimate the utility each would have produced. Then, we consider the utility of the actual and discarded adaptations to determine the ideal one in retrospective, and consequently, the incurred debt. On the other hand, our debt-aware approach observes the debt from a proactive perspective (i.e. before an adaptation takes place); which enables  the Q-learning algorithm to use the learned debts from previous adaptations to forgo immediate rewards in an attempt to pursue a maximization of the utility during the period in which the adaptation is expected to last. Our experiments aim to compare the aggregate utility the approaches can yield and the implication of debt-awareness in utility. We instantiated from the reference system model in figure \ref{fig:model} a scenario with a single debt-aware learning agent.  

\begin{figure}[ht]
\centering
\includegraphics[width=\columnwidth]{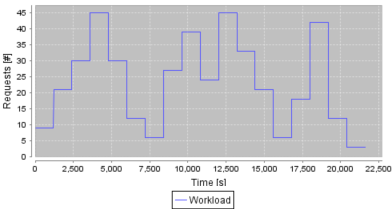}
\caption{Request arrival trace}
\label{fig:arrivaltrace}
\end{figure}

\subsection{Experiment Setup}
We extended CloudSim \cite{calheiros2011cloudsim}, a framework for modelling and simulation of cloud infrastructures and services, and made available extensions in CloudSimEx project \footnote{https://github.com/Cloudslab/CloudSimEx} to support experiments with  both the debt-aware learning and the traditional approach. For the debt-aware learning, we extended Burlap \cite{burlap}, a framework for implementing reinforcement learning solutions, and integrated this extension with CloudSim. Besides the core functionality, we implemented load balancing and horizontal scaling using a single type of virtual machines, where we considered processing capacity expressed in terms of millions of instructions per second (MIPS). Moreover, we use the API  of Apache JMeter \cite{jmeter}, a load and performance tester, to generate a synthetic workload in \textit{Standard Workload Format}, which is the format compatible with CloudSim. The Figure \ref{fig:arrivaltrace} shows a smoothed workload of 6 hours of duration used to represent the arrival rate of requests for our experiments.

General simulation parameters are specified in Table \ref{table:parameters}. Additional specific parameters for the traditional and debt-aware approaches are shown in Tables \ref{table:traditional} and  \ref{table:debt-aware}, respectively. 

\begin{table}[h]
\begin{center}
\caption {Simulation Parameters}  
\begin{tabular}{|c||c|}
 	\hline
    \textbf{Parameter} & \textbf{Value} \\
    \hline
	\hline
	Spin-up time 		& 105s \\
	\hline
	Cool down period 	& 120s \\
	\hline
	Billing cycle 		& Every 5 minutes \\ 
	\hline
	SLA constraint 	& 95\% of jobs handled in less than 2s\\
	\hline
	Price per request 	& \$ 0.0012344 \\
	\hline
	MIPS per request 	& 2 MIPS \\
	\hline
	Penalty per request & \$ 0.002 \\
	\hline  
	VM processing capacity 	& 10 MIPS \\
	\hline
	VM cost 			& \$ 0.01111 per cycle \\
	\hline
\end{tabular}
  \label{table:parameters}
\end{center}
\end{table}

\begin{table}[h]
\begin{center}
\caption {Traditional Approach Simulation Parameters}  
\begin{tabular}{|c||c|}
 	\hline
    \textbf{Parameter} & \textbf{Value} \\
    \hline
	\hline
	Lower CPU threshold & 25\% \\
	\hline
	Upper CPU threshold & 95\% \\
	\hline
	Voting agreement threshold & Relative majority among actions \\
	\hline
\end{tabular}
  \label{table:traditional}
\end{center}
\end{table}

\begin{table}[h]
\begin{center}
\caption {Debt-Aware Approach Simulation Parameters}  
\begin{tabular}{|p{0.39\columnwidth}||p{0.5\columnwidth}|}
 	\hline
    \textbf{Parameter} & \textbf{Value} \\
    \hline
	\hline
	Learning rate \(\alpha\)  & Starts at 1, then decays at 0.1 up to a minimum of 0.1 \\
	\hline
	Discount factor \(\gamma\) 	& 0.99 \\
	\hline
	\(\epsilon\) probability	& 0.1 \\ 
	\hline
	Proportion of VMs with queued requests	& Low (\(<\)15\%) , Medium, High (\(>\)25\%)  \\
	\hline
	Proportion of VMs close to a next billing cycle and without queued requests	& Low (\(<\)33\%) , Medium, High (\(>\)66\%)\\ 
	\hline
\end{tabular}
  \label{table:debt-aware}
\end{center}
\end{table}

We performed the experiments on a laptop that runs Windows 10x64 operating system with 16 GB RAM and Intel Core i7-4500U CPU at 1.8 GHz. Simulations for the traditional approach and the debt-aware learning one took around 89 and 91 seconds, respectively.

\subsection{Results}
We integrated JFreeChart \footnote{http://www.jfree.org/jfreechart/}, a chart library, with CloudSim to draw the charts of resource provisioning, penalties, aggregate utility and elasticity debt over time for each experiment.

For the sake of agility in the discussion, in this subsection, we will indistinguishably refer to the traditional threshold-based rules approach as the \textit{traditional approach}, whereas we will use the term \textit{debt-aware approach} in reference to the debt-aware reinforcement learning approach.

\begin{figure}[ht]
\centering
\includegraphics[width=\columnwidth]{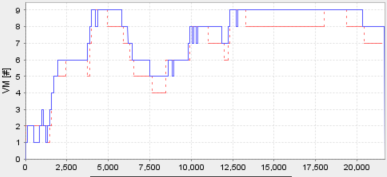}
\caption{Resource provisioning in the debt-aware learning approach}
\label{fig:learning-provisioning}
\end{figure}

\begin{figure}[ht]
\centering
\includegraphics[width=\columnwidth]{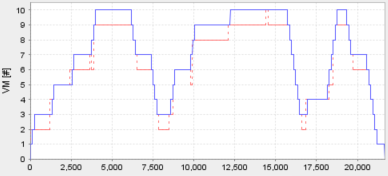}
\caption{Resource provisioning in the traditional approach}
\label{fig:rule-provisioning}
\end{figure}

We compare resource provisioning graphs obtained from both approaches. Figures \ref{fig:learning-provisioning} and \ref{fig:rule-provisioning} illustrate the actual resource provisioning overlaid with the ideal one corresponding to the debt-aware and traditional  approaches, respectively. The costs of VMs provisioned by the debt-aware approach were \$5.98; whereas the traditional approach incurred in \$ 5.75. The debt-aware approach shows to be more unstable at the beginning, as a result of the initial explorations. However, much later appears to be more stable than the traditional one. This stability increased costs related to deployed VMs; but it produced a lower number of penalties, as appreciated in the penalties over time charts.

\begin{figure}[ht]
\centering
\includegraphics[width=\columnwidth]{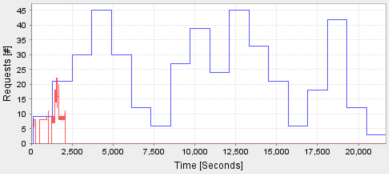}
\caption{Penalties over time in the debt-aware learning approach}
\label{fig:learning-penalties}
\end{figure}

\begin{figure}[ht]
\centering
\includegraphics[width=\columnwidth]{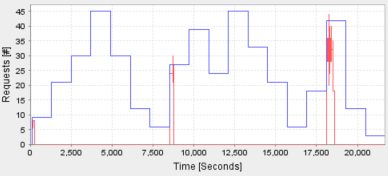}
\caption{Penalties over time in the traditional approach}
\label{fig:rule-penalties}
\end{figure}

Figures \ref{fig:learning-penalties} and \ref{fig:rule-penalties} depict the penalties over time from both approaches overlaid with the submitted requests in the workload. The debt-aware approach had a 2.86 \% of failed requests; whereas the traditional failed a 4.17 \% of submitted requests. Therefore, higher costs of resource provisioning of the debt-aware approach were compensated by a lower number of SLA violations. Additionally, as we can notice, the penalties in the debt-aware approach are concentrated at the beginning of the execution, where a higher exploration of debts takes place.

\begin{figure}[ht]
\centering
\includegraphics[width=\columnwidth]{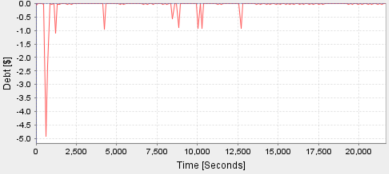}
\caption{Debt over time in the debt-aware learning approach}
\label{fig:learning-debt}
\end{figure}

\begin{figure}[ht]
\centering
\includegraphics[width=\columnwidth]{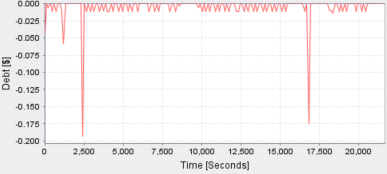}
\caption{Debt over time in the traditional approach}
\label{fig:rule-debt}
\end{figure}

According to expectations, the experiment shows that elasticity adaptations to adjust resource provisioning in the deb-aware approach, depicted in Figure \ref{fig:learning-debt}, incurs in higher debts in exploration stages; but during exploitation stages, debts tend to be minimized. On the other hand, Figure \ref{fig:rule-debt} shows that adaptations in the traditional approach incurs in elasticity debts without emerging a specific pattern over time, which may lead to unintentional debts that repeatedly affect the aggregate utility. 

\begin{figure}[ht]
\centering
\includegraphics[width=\columnwidth]{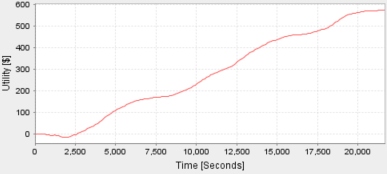}
\caption{Aggregate utility in the debt-aware learning approach}
\label{fig:learning-utility}
\end{figure}

\begin{figure}[ht]
\centering
\includegraphics[width=\columnwidth]{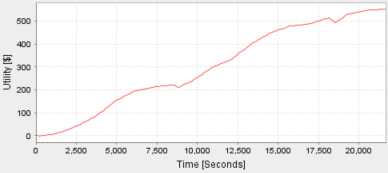}
\caption{Aggregate utility in the traditional approach}
\label{fig:rule-utility}
\end{figure}

The aggregate utility over time achieved by both approaches are depicted in Figures \ref{fig:learning-utility} and \ref{fig:rule-utility}. The aggregate utility of the debt-aware approach reached \$ 573.55 at the end of the workload processing; whereas the traditional approach yielded \$ 552.13. This implies that elasticity debt considerations increased the overall utility achieved by the SaaS provider in a 4 \%. Additionally, we can observe that the aggregate utility in the debt-aware approach grows more smoothly than the traditional; which presents two drops that hurt the utility at time 8750 and 18500 seconds, approximately.

\subsection{Threats to Validity}
We carried out the evaluation of our approach through a simulation that resembles a cloud environment. However, we built our simulation tool on CloudSim, JMeter and Burlap, which are the most widely extended frameworks for simulating cloud environments, generating synthetic workload and implementing reinforcement learning experiments, respectively. Additionally, our controlled environment facilitates a faster experimentation with diverse scenarios and different IaaS providers.

For the sake of simplicity, we considered a SLA with only one quality of service attribute: response time. Nonetheless, the model is extensible to multiple attributes (e.g. availability, reliability). 

The overhead of the debt-aware approach may be affected if more variables are included to define a state. However, our experiments show that even with only two variables, the approach is able to yield a better outcome than a traditional one.  

The instantiation has considered a single debt-aware learning agent; considering multiple learners can introduce several degrees of freedom in the experimentation due to the complexity that can arrive from learning agents communicating. An extensive experimental study covering the above issues is worthy separate systematic study. 

\section{Related Work}
Technical debt community has applied the metaphor in a wide range of decision-making process under uncertainty such as software maintenance and evolution \cite{kruchten2012technical}, architectural design \cite{li2014architectural}, cloud service selection \cite{alzaghoul2013economics}, software testing, sustainability design among others \cite{li2015systematic}. It has been used as a way to identify, measure and monitor a decision that trades off a quality compliance concern against an economics concern. Furthermore, the metaphor has shown to be effective to raise the visibility of the impact on utility of a suboptimal decision if a change materialises. For example, Li et. al. \cite{li2014architectural} evaluated architectural decisions from a value-oriented perspective and used the debt to monetise the gap between an optimal and suboptimal architecture when a change scenario occurs. Also, Alzaghoul et. al.  \cite{alzaghoul2013economics} extended the metaphor into cloud service selection to adopt a service substitution that is aware of the potential debt introduced in the composition by each candidate service and makes a decision based on the potential of the selected service to clear the debt when the change scenario materialise. However, none of these works addresses the problem of automating the learning of technical debts. To the best of our knowledge, we are the first to propose an autonomous management of technical debts based on learning and, different from previous works, we are revisiting the metaphor to support run-time management of debts and value creation in self-adaptive and self-management contexts such as cloud elasticity.

Reinforcement learning has already been used as an underlying technique for elasticity management \cite{lorido2014review}. For instance, Barret et. al. \cite{barrett2013applying} designed a parallel Q-learning approach to build an elasticity manager based on a multi-agent system, where each virtual resource is an agent that makes its decisions depending on the load of incoming requests, experienced penalties and deploying costs. However, the approach ignores the debt introduced to the utility of an elasticity adaptation when elasticity determinants such as billing cycles or spin-up times are neglected. Jamshidi et. al. \cite{jamshidi2016managing, jamshidi2014autonomic} built a fuzzy control based reinforcement learning approach for autonomous elasticity management that modifies fuzzy elasticity rules for resource provisioning at run-time. However, this work is focused on tuning and improving fuzzy rules to reduce user-dependency in elasticity management. In contrast to prior works, we designed a reinforcement learning approach that considers state variables related to both economics and performance aspects of cloud elasticity, in order to achieve a management that proactively uses this autonomous learning of technical debts in resource adaptations to estimate the conditions where these debts will potentially pay off. 



\section{Conclusions and Future Work}
We proposed an autonomous elasticity management approach intended to make adaptations that are aware of the unavoidable elasticity imperfections due to the dynamic trade-offs between economics and performance of elasticity adaptations in the cloud. Our approach performs a dynamic reinforcement learning of the gaps between the ideal and actual resource provisioning over time.  We are the first to propose an elasticity decision-making analysis that integrates the strategic decision-making achieved through reinforcement learning techniques, and the value oriented perspective promoted by the technical debt metaphor in changing environments. Simulation results show that a reinforcement learning of dynamic technical debts in resource provisioning can achieve a higher aggregate utility for a SaaS provider. Moreover, the underlying foundations of our dynamic technical debt approach are applicable in other self-adaptive and self-management contexts, where decisions with a trade-off analysis can be strategically taken and aimed at long-term rewards.

In our ongoing research, we are identifying the properties of elasticity debts and linking them to the sources of uncertainty in the cloud. Additionally, we are introducing a technical debt-oriented perspective for multi-tenant applications hosted in inter-clouds architectures. 


\section*{Acknowledgment}

We thank Tao Chen for his helpful comments on the paper.



%
%

%



\end{document}